\documentclass[AMA,Times1COL]{WileyNJDv5} 
\usepackage{graphicx} 

\usepackage[nomarkers,nolists]{endfloat}

\def\T{{ \mathrm{\scriptscriptstyle T} }}

\articletype{Journal Article}%

\received{Date Month Year}
\revised{Date Month Year}
\accepted{Date Month Year}
\journal{Journal}
\volume{00}
\copyyear{2025}
\startpage{1}

\begin{document}

\title{Regression for Left-Truncated and Right-Censored Data: A Semiparametric Sieve Likelihood Approach}

\author[1]{Spencer Matthews}

\author[2]{Bin Nan}


\authormark{MATTHEWS and NAN}
\titlemark{REGRESSION FOR LEFT-TRUNCATED AND RIGHT-CENSORED DATA}

\address[]{\orgdiv{Department of Statistics}, \orgname{University of California, Irvine}, \orgaddress{\state{California}, \country{U.S.A.}}}



\corres{Corresponding author Bin Nan. \email{nanb@uci.edu}}



\abstract[Abstract]{Cohort studies of the onset of a disease often encounter left-truncation on the event time of interest in addition to right-censoring due to variable enrollment times of study participants. Analysis of such event time data can be biased if left-truncation is not handled properly. We propose a semiparametric sieve likelihood approach for fitting a linear regression model to data where the response variable is subject to both left-truncation and right-censoring.
We show that the estimators of regression coefficients are consistent, asymptotically normal and semiparametrically efficient. Extensive simulation studies show the effectiveness of the method across a wide variety of error distributions. We further illustrate the method by analyzing a dataset from The 90+ Study for aging and dementia.}



\maketitle

\renewcommand\thefootnote{}

\renewcommand\thefootnote{\fnsymbol{footnote}}
\setcounter{footnote}{1}

\section{Introduction}

\label{sec:intro}

There has been significant research effort in understanding age-related diseases, such as Alzheimer's disease and dementia, in recent decades.
A key method in understanding these diseases is to design a prospective cohort study, where study participants are recruited and followed, with data collected at regular intervals. Oftentimes the ultimate outcome of interest is the time at which the disease occurs.
Such studies typically require that participants who enroll in the study meet an age requirement while being disease free; for instance, The 90+ Study only enrolled dementia-free individuals at age 90 or older \citep{corrada_dementia_2010}.
This method of participant enrollment induces left-truncation into the resulting dataset. 

Left-truncation occurs when individuals who have already developed the condition of interest prior to the commencement of the study are not observed or excluded from the analysis. Simply ignoring left-truncation introduces bias as individuals who developed the disease earlier are not represented, potentially skewing the findings towards a healthier or longer-lived population. Appropriately accounting for left-truncation is crucial in ensuring the validity of study results, particularly in longitudinal research investigating age-related diseases such as Alzheimer's disease and dementia. 

Another common feature of event time data is right-censoring, which occurs when some participants in the study do not experience the event of interest within the specified study time period, or are lost to follow-up before the event can be recorded. As a result, the exact time of the event for such an individual remains unknown, and the available information is limited to the fact that the event has not occurred up to the last observed time.

While fitting parametric models or semiparametric hazard regression models, e.g., the Cox model, to data subject to left-truncation and right-censoring has been studied extensively, the case of fitting a semiparametric accelerated failure time model to such data is underdeveloped. 
Let $\tilde Y$ be the underlying event time and define $Y^* = \xi(\tilde Y)$ where $\xi(\cdot)$ is a well-defined continuous and monotone transformation. 
Suppose $\tilde Y$ can be left-truncated by $\tilde T$ and right-censored by $\tilde C$. Let $T = \xi(\tilde T)$ and $C = \xi(\tilde C)$, then $Y^*$ can be left-truncated by $T$ and right-censored by $C$.
Define $Y = Y^*\wedge C$, where $a \wedge b = \min{(a,b)}$, and $\Delta = {1}(Y^* \leq C)$, where $1(\cdot)$ is an indicator function. We also define $a \vee b = \max{(a,b)}$.
Consider the following linear regression model given covariates $X$:
\begin{equation}
    Y^* = X^\T \beta + e,
    \label{eqn:aft}
\end{equation}
where $\beta \in {R}^d$, $d$ is a fixed integer, and the error $e$ follows some unknown distribution $F_e$ with density $f_e$.  Thus model (\ref{eqn:aft}) is a semiparametric model.
If $\xi(\cdot) = \log(\cdot)$, then this becomes the so-called accelerated failure time model as in \citet{kalbfleisch_statistical_2002}, Chapter 7. We extend the terminology to model (\ref{eqn:aft}) with any known transformation $\xi$.  

One method for fitting such a semiparametric model is rank regression. \citet{adichie_estimates_1967} was among the first who used residual rank statistics for fitting regression parameters, and \citet{bhattacharya_nonparametric_1983} extended the approach to left-truncated data in their work in astronomy.
\citet{lai_rank_1991, lai_linear_1992, lai_missing_1994} used these ideas in exploring rank statistics with left-truncated and right-censored data as they studied the behavior and asymptotic properties of such a regression.
\citet{kim_efficient_2000} pointed out various issues with practical implementations of rank regression methods.
In an attempt to overcome some of these issues, they introduced an adaptive M-estimation method using efficient scores for censored and truncated regression models in which they used splines to approximate their estimating equations.
While they showed their method to have nice theoretical properties, the implementation is dense and no code package is readily available.

We consider a sieve likelihood approach for fitting model (\ref{eqn:aft}) to left-truncated and right-censored data where the unknown infinite-dimensional nuisance parameter, the logarithm of the hazard function of $e$ in our case, is estimated using splines.
Our method builds on the work of \citet{ding_sieve_2011} for bundled parameters where only right-censoring was considered.
This approach allows for a semiparametric likelihood estimation for a frequently encountered problem in practice, which not only provides a simple numerical implementation but also yields desirable asymptotic properties for the regression coefficient estimates, namely asymptotical normality and semiparametric efficiency.
We demonstrate this approach through a simulation study and an application to The 90+ Study for dementia among the oldest old \citep{corrada_dementia_2010}.

\section{Sieve Likelihood Approach}
\label{sec:likelihood}

A sieve approach approximates the true unknown infinite-dimensional parameter with a sequence of increasingly complex parametric models. In our case, we propose to use a sequence of B-spline basis expansions to approximate the logarithm of the unknown hazard function of $e$ in model (\ref{eqn:aft}). 

Denote the hazard function of $e$ by $\lambda_e$, and let $g = \log\lambda_e$. 
For a finite interval $[a,b]$ and $K$ knots placed within the interval: $a < \xi_{1} < \ldots < \xi_{K} < b$, we approximate $g$ by 
$$g(s) = \log \lambda_e(s) \approx \sum_{k=1}^K \gamma_kB_k(s),$$
where $B_k, k=1, \dots, K,$ are cubic B-spline basis functions. We chose cubic B-splines so that the second derivative is continuous, but other splines may be considered as well. For further discussion see \citet{de_boor_practical_2001}, Chapter 9.

Since $e = Y^* - X^\T\beta$, $g$ is a composite function of $\beta$. Thus $(\beta, g)$ form a pair of bundled parameters \citep{ding_sieve_2011}. Denote $\epsilon_{\beta, i} = Y_i - X_i^\T \beta$ and $\tau_{\beta, i} = T_i - X_i^\T \beta$. Assume $Y^*$ is independent of $(C, T)$ given $X$. Then for independent and identically distributed observations $\{(Y_i, \Delta_i, X_i, T_i): i=1, \dots,n\}$, we have the following likelihood function:
\begin{equation*}
     \prod_{i=1}^n\left[\exp\{g(\epsilon_{\beta, i})\}\right]^{\Delta_i} \exp\left[-\int_{\tau_{\beta,i}}^{\epsilon_{\beta,i}} \exp\{g(u)\}du\right],
\end{equation*}
which is then approximated by the sieve likelihood:
\begin{equation*}
    L_n(\beta, \gamma) = \prod_{i=1}^n\left[\exp\left\{\sum_{k=1}^K \gamma_kB_k(\epsilon_{\beta, i})\right\}\right]^{\Delta_i} \exp\left[-\int_{\tau_{\beta,i}}^{\epsilon_{\beta,i}} \exp\left\{\sum_{k=1}^K \gamma_kB_k(u)\right\}du\right].
\end{equation*}
See the Appendix for the detailed construction of the density function of $(Y, \Delta, X, T)$ given $Y > T$ which leads to the above likelihood function for data with left-truncation and right-censoring. 
Then we obtain the following sieve log-likelihood function:
\begin{equation}
    \label{eqn:loglik}
    \ell_n(\beta, \gamma) = \sum_{i=1}^n \left(\Delta_i\sum_{k=1}^K \gamma_k B_k\left(\epsilon_{\beta, i}\right) - \int_{\tau_{\beta, i}}^{\epsilon_{\beta, i}}\exp\left\{\sum_{k=1}^K \gamma_k B_k(s)\right\}ds\right).
\end{equation}

Denote $\dot{B}_k(s) = d B_k(s)/ds$ and $\ddot{B}_k = d^2B_k(s) / ds^2$. 
Taking partial derivatives of the above sieve log-likelihood function (\ref{eqn:loglik}), we arrive at the following sieve score functions:
\begin{equation}
    S_n(\beta, \gamma) = \left(
        \frac{\partial\ell_n(\beta, \gamma) }{\partial \beta},
        \frac{\partial\ell_n(\beta, \gamma) }{\partial \gamma} \right)^\T,
    \label{eqn:score}
\end{equation}
with
\begin{equation}
    \label{eqn:score1}
    \begin{aligned}
    \frac{\partial\ell_n(\beta, \gamma)}{\partial \beta} &= \sum_{i=1}^nX_i^\T\left( -\Delta_i \sum_{k=1}^K\gamma_k \dot{B}_k\left(\epsilon_{\beta, i}\right) + \exp\left\{\sum_{k=1}^K \gamma_k B_k\left(\epsilon_{\beta, i}\right)\right\} \right. \\
    &\quad\quad\quad\quad \left.- \exp\left\{\sum_{k=1}^K \gamma_k B_k\left(\tau_{\beta, i}\right)\right\}\right)
    \end{aligned}
\end{equation}
and
\begin{equation}
    \label{eqn:score2}
    \frac{\partial\ell_n(\beta, \gamma)}{\partial \gamma_k} = \sum_{i=1}^n \left(\Delta_i B_k\left(\epsilon_{\beta, i}\right) - \int_{\tau_{\beta, i}}^{\epsilon_{\beta, i}} B_k(s)\exp \left\{\sum_{k=1}^K \gamma_k B_k(s)\right\}ds\right).
\end{equation}

Furthermore, the observed sieve information matrix is
\begin{equation}
    \label{eqn:info}
    I_n(\beta, \gamma) = \begin{pmatrix}
        \frac{\partial^2\ell_n(\beta, \gamma)}{\partial\beta\partial\beta^\T}  & \frac{\partial^2\ell_n(\beta, \gamma)}{\partial\beta  \partial\gamma^\T} \\
        \frac{\partial^2\ell_n(\beta, \gamma)}{\partial\gamma \partial\beta^\T} & \frac{\partial^2\ell_n(\beta, \gamma)}{\partial\gamma\partial\gamma^\T}
    \end{pmatrix},
\end{equation}
with
\begin{eqnarray*}
    \frac{\partial^2\ell_n(\beta, \gamma)}{\partial\beta\partial\beta^\T} &=& \sum_{i=1}^n X_iX_i^\T\left(\Delta_i\sum_{k=1}^K \gamma_k \ddot{B}_k\left(\epsilon_{\beta, i}\right) -  \exp\left\{\sum_{k=1}^K \gamma_k B_k\left(\epsilon_{\beta, i}\right)\right\} \sum_{k=1}^K\gamma_k \dot{B}_k(\epsilon_{\beta, i})\right. \\
    &&\quad\quad\quad\quad+ \left.\exp\left\{\sum_{k=1}^K \gamma_k B_k\left(\tau_{\beta, i}\right)\right\} \sum_{k=1}^K\gamma_k \dot{B}_k\left(\tau_{\beta, i}\right)\right),
    \\
        \frac{\partial^2\ell_n(\beta, \gamma)}{\partial\beta  \partial\gamma_k} &=& \sum_{i=1}^n X_i^\T\left(-\Delta_i \dot{B}_k\left(\epsilon_{\beta, i}\right) + B_k\left(\epsilon_{\beta, i}\right)\exp\left\{\sum_{k=1}^K \gamma_k B_k\left(\epsilon_{\beta, i}\right)\right\} \right. \\
        &&\quad\quad\quad\quad \left.- B_k\left(\tau_{\beta, i}\right)\exp\left\{\sum_{k=1}^K \gamma_k B_k\left(\tau_{\beta, i}\right)\right\}\right),
    \\
    \frac{\partial^2\ell_n(\beta, \gamma)}{\partial\gamma_k  \partial\gamma_l} &=& \sum_{i=1}^n \left(-\int_{\tau_{\beta, i}}^{\epsilon_{\beta, i}} B_k(s) B_l(s)\exp\left\{\sum_{m=1}^K \gamma_m B_m(s)\right\} ds\right).
\end{eqnarray*}

We consider a gradient search algorithm by setting (\ref{eqn:score1}) and (\ref{eqn:score2}) equal to zero and solving for $\beta$ and $\gamma$, which leads to the sieve maximum likelihood estimates $(\hat\beta_n, \hat\gamma_n)$. Similar to \citet{ding_sieve_2011} we can see that, under the regularity conditions given in the next section, the sieve log-likelihood function (\ref{eqn:loglik}) is locally concave in a neighborhood of true parameters $(\beta_0, g_0)$. Thus, we start the search algorithm from multiple randomly selected initial values and set $(\hat\beta_n, \hat\gamma_n)$ to be the converging result that gives the maximum value of the sieve log-likelihood (\ref{eqn:loglik}). Specifically in our numerical examples, we generate 10 random initial values each obtained by adding a vector of random noises to the naive ordinary least squares estimator of $\beta$ obtained from complete observations, where the standard deviation of each random noise is chosen to be three times the standard error of the corresponding estimated coefficient from the naive model. We multiply the standard error by three to avoid getting caught too close to the naive estimate for all random starts. We sample initial $\gamma$ values from the standard normal distribution. To avoid singular information matrix during search iterations, we only use diagonal elements of the sieve information matrix (\ref{eqn:info}), with the final full information matrix evaluated at $(\hat\beta_n, \hat\gamma_n)$. The method works well in our simulation studies presented in Section \ref{sec:simulations}.

\section{Asymptotic Results}
\label{sec:asymptotic}

Define $\epsilon_0 = Y - X^\T\beta_0$ and $\tau_0 = T - X^\T\beta_0$, where $\beta_0$ is the true parameter. Let $\lambda_0$ be the hazard function of $e_0 = Y^* - X^\T\beta_0$ and $\Lambda_0$ be the corresponding cumulative hazard function. Following \citet{lai_rank_1991} we assume $e_0$ is independent of $(C,T,X)$, which implies that $Y^*$ is independent of  $(C,T)$ given $X$, an assumption we need for obtaining the joint density function of $(Y, \Delta, X,T)$ given $Y>T$ in the Appendix. To obtain desirable asymptotic properties of the proposed estimator, 
we introduce the same regularity conditions in \citet{ding_sieve_2011} with a few additional assumptions for the left-truncation.

\vspace{1em}

\noindent(A.1) The true parameter $\beta_0$ belongs to the interior of a compact set $\mathcal{B} \subseteq {R}^d$.

\noindent(A.2) (a) The covariate $X$ takes values in a bounded subset $\mathcal{X} \subseteq {R}^d$; (b) $E(XX^\T)$ is non-singular.

\noindent(A.3) (a) There is a constant $b < \infty$ such that, for some constant $\delta$, ${\rm pr}(\epsilon_0 > b \mid X) \geq \delta > 0$ almost surely with respect to the probability measure of $X$. This implies that $\Lambda_{0}(b) \leq -\log\delta < \infty$;
(b) We also assume ${\rm pr}(\tau_0 < b \mid X) \geq 1 - \delta_1$ for some constant $\delta_1 \in (0,1)$ almost surely, and ${\rm pr}(\tau_0 > \epsilon_0 \mid X) \leq \delta_2$ for some constant $\delta_2 \in [0,1)$ almost surely. 

\noindent(A.4) The error $e_0$'s density $f_0$ and its derivative $\dot f_0$ are bounded and 
\[
\int \left\{\dot f_0(s) / f_0(s)\right\}^2f_0(s)ds < \infty.
\]

\noindent(A.5) (a) The conditional density of $C$ given $(T,X)$ and its derivative are uniformly bounded for all possible values of $T \in\mathcal{T} \subseteq {R}$ and $X\in\mathcal{X}$, that is,
\[
\sup_{t\in\mathcal{T},x\in\mathcal{X}} f_{C\mid T, X}(y \mid t,x) \leq K_1, \quad\quad \sup_{t \in\mathcal{T}, x\in\mathcal{X}}\left|\dot f_{C\mid T, X}(y\mid t, x)\right| \leq K_2
\]
for all $y \in \mathcal{Y} \subseteq {R}$ with some constants $K_1, K_2 > 0$; 
(b) Similarly, the conditional density of $T$ given $X$ and its derivative are uniformly bounded for all possible values of $X\in\mathcal{X}$, that is,
\[
\sup_{x\in\mathcal{X}} f_{T\mid X}(t \mid  x) \leq K_3, \quad\quad \sup_{x\in\mathcal{X}} \left|\dot f_{T\mid X}(t\mid  x)\right| \leq K_4
\]
for all $t \in \mathcal{T}$ with some constants $K_3, K_4 >0$; (c) Both $f_{C\mid T,X}$ and  $f_{T,X\mid Y>T}$ are free of parameters $\beta$ and $\lambda_e$.

\noindent(A.6) Let $\mathcal{G}^p$ denote the collection of bounded functions $g$ on $[a,b]$ with bounded derivatives $g^{(j)}, j = 1, \ldots, k$. The $k$th derivative $g^{(k)}$ satisfies the following Lipschitz continuity condition:
\[
\left|g^{(k)}(s) - g^{(k)}(t)\right| \leq  L|s-t|^m \quad\quad \text{for } s,t \in [a,b],
\]
where $k$ is a positive integer and $m \in (0,1]$ such that $p = k+m \geq 3$, $L<\infty$ is an unknown constant,  $a=\inf_{y,x}(y-x^\T\beta_0) \vee \inf_{t,x}(t-x^\T\beta_0) > -\infty$ and $b$ is defined in condition (A.3). The true log hazard function of the error $e_0$, $g_0(\cdot) = \log\lambda_0(\cdot)$, belongs to $\mathcal{G}^p$.

\noindent(A.7) For some $\eta \in (0,1)$, $u^\T\text{Var}(X \mid \epsilon_0)u \geq \eta u^\T E(XX^\T\mid \epsilon_0)u$ almost surely for all $u\in {R}^d$.

\vspace{1em}

The modified regularity conditions to those given in Section 3 of \citet{ding_sieve_2011} are in (A.3)(b) and (A.5). 
Condition (A.3)(b) guarantees that we observe left-truncated data almost surely.  
Condition (A.5)(a) is a natural extension of the condition in \citet{ding_sieve_2011} for the conditional density function of $C$. Condition (A.5)(b) plays a similar role as (A.5)(a). 
Condition (A.5)(c) contains a noninformative censoring assumption for right-censored data and a similar but less stringent noninformative truncation assumption made by \citet{lai_asymptotically_1992} for left-truncated data. Condition (A.5)(c) simplifies the likelihood function to the desirable form we use in this article. 
When $\sup\{t \in \mathcal{T}\} \le \inf\{y\in\mathcal{Y}\}$, left-truncation does not play any role so we only deal with right-censored data and the problem reduces to that in \citet{ding_sieve_2011}. On the other hand, when ${\rm pr}(Y^*\le C)=1$, the problem reduces to left-truncation only.

Consider the collection of functions
\[
\mathcal{H}^p = \left\{\zeta(\cdot, \beta)\colon \zeta(s, x, \beta) = g(\psi(s, x, \beta)), g \in \mathcal{G}^p, s\in [a,b], x\in\mathcal{X}, \beta \in \mathcal{B}\right\},
\]
where $\psi(s, x, \beta) = s - x^\T(\beta - \beta_0)$ and $\mathcal{G}^p$ is defined in Condition (A.6).
Here $\zeta$ is a composite function of $g$ composed with $\psi$. Note that $\zeta(s, x, \beta_0) = g(s)$.
Then for $\zeta(\cdot, \beta) \in \mathcal{H}^p$ we define the following norm:
\begin{equation*}
    \lVert \zeta(\cdot, \beta) \rVert_2 = \left[\int_{\mathcal{X}}\int_a^b \left\{g\left(s - x^\T(\beta - \beta_0)\right)\right\}^2 d\Lambda_0(s) dF_X(x)\right]^{1/2}.
\end{equation*}

For any $\theta_1 = (\beta_1, \zeta_1(\cdot, \beta_1))$ and $\theta_2 = (\beta_2, \zeta_2(\cdot, \beta_2))$ in the space of $\Theta^p = \mathcal{B} \times \mathcal{H}^p$, define the following distance:
\begin{equation}
    d(\theta_1, \theta_2) = \left\{|\beta_1 - \beta_2|^2 + \lVert\zeta_1(\cdot, \beta_1) - \zeta_2(\cdot, \beta_2)\rVert_2^2\right\}^{1/2}.
    \label{eqn:distance}
\end{equation}
This is the same distance used by \citet{ding_sieve_2011} in their treatment of the right-censored case.

Let $\mathcal{G}_n^p$ be the space of polynomial splines of order $p \geq 1$. For details on this space see \citet{schumaker_spline_2007}, Chapter 4. Denote 
\[
\mathcal{H}^p_n = \left\{\zeta(\cdot, \beta)\colon \zeta(s, x, \beta) = g(\psi(s, x, \beta)), g \in \mathcal{G}^p_n, s\in [a,b], x\in\mathcal{X}, \beta \in \mathcal{B}\right\}
\]
and $\Theta_n^p = \mathcal{B} \times \mathcal{H}_n^p$. Clearly, $\mathcal{H}_n^p \subseteq \mathcal{H}_{n+1}^p \subseteq \cdots \subseteq \mathcal{H}^p$ for all $n \geq 1$. The sieve estimator $\hat\theta_n = (\hat\beta_n, \hat\zeta_n(\cdot, \hat\beta_n))$, where $\hat\zeta_n(s, x, \hat\beta_n) = \hat g_n(s - x^\T(\hat\beta_n - \beta_0))$, is the maximizer of the log-likelihood over the sieve space $\Theta_n^p$. The following theorem gives the convergence rate of the proposed estimator $\hat\theta_n$ to the true parameter $\theta_0 = (\beta_0, \zeta_0(\cdot, \beta_0)) = (\beta_0, g_0)$.

\begin{theorem}
\label{them:consistency}
    Let $K_n$ be the number of internal knots in the spline space and $K_n = O(n^\nu)$, where $\nu$ satisfies $1/\{2(1+p)\} < \nu < 1/(2p)$ with $p$ being the smoothness parameter defined in condition (A.6). Suppose conditions (A.1)--(A.7) hold. Then
    \[
    d(\hat\theta_n, \theta_0) = O_p\left\{n^{-\min(p\nu, (1 - \nu) / 2)}\right\},
    \]
    where $d(\cdot, \cdot)$ is defined in (\ref{eqn:distance}).
\end{theorem}

\begin{remark}
    The sieve space $\mathcal{G}_n^p$ does not have to be restricted to the B-spline space. It can be any sieve space as long as the estimator $\hat\theta_n \in \mathcal{B} \times \mathcal{H}_n^p$ satisfies the conditions of Theorem 1 in \citet{shen_convergence_1994}. Our choice of the cubic B-spline space is primarily motivated by its simplicity of numerical implementation, which is a tremendous advantage of the proposed approach over existing numerical methods for fitting accelerated failure time models for data subject to right-censoring and left-truncation.
\end{remark}

The proof of Theorem \ref{them:consistency} follows that of \citet{ding_supplement_2011} by verifying the conditions of \citet{shen_convergence_1994}. The main difference is controlling the additional complication brought by $\tau_\beta= T-X^\T\beta$ in the likelihood function, which is managed by Condition (A.3)(b) as well as desirable properties of linear and indicator  functions.  
Theorem \ref{them:consistency} implies that if $\nu = 1 / (1 + 2p)$, $d(\hat\theta_n, \theta_0) = O_p(n^{-p/(1+2p)})$ which is the optimal convergence rate in the nonparametric regression setting. 
Although the overall convergence rate is slower than $n^{-1/2}$, the next theorem states that the proposed estimator of the regression parameter is still asymptotically normal and semiparametrically efficient.

\begin{theorem}
    \label{them:normality}
    Given the following efficient score function obtained by \citet{lai_asymptotically_1992} for the censored and truncated linear model:
    \[
    l^*_{\beta_0}(Y, \Delta, X, T) = \int \{X - E[X \mid \epsilon_0 \geq s, \tau_0 \leq s]\}\left\{\frac{-\dot\lambda_0}{\lambda_0}(s)\right\}dM(s),
    \]
    where
    \[
    M(s) = \Delta 1(\epsilon_0 \leq s)1(\epsilon_0 \geq \tau_0) - \int_{-\infty}^s 1(\epsilon_0 \geq u)1(\tau_0 \leq u) \lambda_0(u)du
    \]
    is the failure counting process martingale, suppose that the conditions in Theorem \ref{them:consistency} hold and $I(\beta_0) = P\{l^*_{\beta_0}(Y, \Delta, X, T)^{\otimes 2}\}$ is nonsingular. Then
    \[
    n^{1/2}(\hat\beta_n - \beta_0) = n^{-1/2} I^{-1}(\beta_0) \sum_{i=1}^n l^*_{\beta_0}(Y_i, \Delta_i, X_i, T_i) + o_p(1) \to N(0, I^{-1}(\beta_0))
    \]
    in distribution.
\end{theorem}

The proof of Theorem \ref{them:normality} closely follows the work of \citet{ding_sieve_2011, ding_supplement_2011} and \citet{zhao_sieve_2017} by checking Conditions (A1) through (A6) of \citet{ding_sieve_2011}. Verifying their Condition (A3) requires searching for the least-favorable submodel, which leads to the same efficient score function found by \citet{lai_asymptotically_1992}. The main difference to \citet{ding_sieve_2011} is again the inclusion of the left-truncation indicator, which has minimum impact under the additional regularity conditions (A.3)(b) and (A.5)(b) and desirable properties of indicator and linear functions.  

The following theorem provides the consistency result of the variance estimator based on the efficient score function given in Theorem \ref{them:normality}. Denote $Pf = \int f(y,\delta,x,t)dP(y,\delta,x,t)$ and ${P}_nf = n^{-1}\sum_{i=1}^n f(Y_i,\Delta_i,X_i,T_i)$, where $P$ is a probability measure,
and ${P}_n$ is an empirical probability measure.

\begin{theorem}
    \label{them:var_con}
    Suppose the conditions in Theorem \ref{them:normality} hold. Denote 
    \[
    \hat{l}^*_{\hat\beta_n}(Y, \Delta, X, T) = \int \{X - \bar{X}(s; \hat\beta_n)\}\{-\dot{\hat g}_n(s)\}d\hat M(s),
    \]
    where
    \[
    \bar{X}(s; \hat\beta_n) = \frac{{P}_n\{X 1(Y - X^\T\hat\beta_n \geq s)1(T - X^\T\hat\beta_n \leq s)\}}{{P}_n\{1(Y - X^\T\hat\beta_n \geq s)1(T - X^\T\hat\beta_n \leq s)\}}
    \]
    and
    \begin{align*}
    \hat M(s) &= \Delta 1(T - X^\T\hat\beta_n \leq Y - X^\T\hat\beta_n \leq s) \\
    &\quad\quad\quad\quad - \int_{-\infty}^s 1(Y - X^\T\hat\beta_n \geq u)1(T - X^\T\hat\beta_n \leq u) \exp\{\hat g_n(u)\} du.
    \end{align*}
    Then $${P}_n\left\{\hat{l}^*_{\hat\beta_n}(Y, \Delta, X, T)^{\otimes 2}\right\} \to P\left\{l^*_{\beta_0}(Y, \Delta, X, T)^{\otimes 2}\right\}$$ in probability.
\end{theorem}

In Theorem \ref{them:var_con}, $\bar{X}(s, \hat\beta_n)$ estimates $E[X \mid \epsilon_0 \geq s, \tau_0 \leq s]$ from Theorem \ref{them:normality}. Once again, the proof of Theorem \ref{them:var_con} follows  \citet{ding_supplement_2011} with a major difference of including the truncation indicator, which can be easily dealt with. Due to a large duplication of \citet{ding_sieve_2011} in the proofs of above theorems, we omit all the details in this article.  

\section{A Simulation Study}
\label{sec:simulations}

\subsection{Simulation Setup}

In order to assess the validity of the proposed methodology, we perform an extensive simulation study with the following setup.
We simulate datasets from the following linear model:

\begin{equation}
    \label{eqn:model}
    Y = \beta_1X_1 + \beta_2X_2 + e
\end{equation}
with $\beta_1=\beta_2=1$. We generate continuous covariate $X_1$  from a Uniform(-3, 3) distribution,  binary covariate $X_2$ from a Bernoulli(0.5) distribution, truncation time $T$  from a Uniform(-6, 1) distribution, and  censoring times $C$  from a Uniform(1, 7) distribution.

The error term comes from one of four distributions: (a) standard normal distribution, (b) standard extreme value distribution, or Gumbel distribution, (c) a 50\% even mixture of standard normal and $N(0, 3^2)$ distributions, and (d) a 50\% even mixture of standard normal and $N(-1, 0.5^2)$ distributions. Table \ref{tbl:sim_ops} shows the
operating characteristics of such simulation setups.
In addition we also note in the table the number of internal knots used to estimate the nuisance parameter for each setup, which is determined  via cross-validation.

\begin{table}
\centering
	\caption{Simulation operating characteristics under different error distributions}
{	\begin{tabular}{lccccc} 

    \toprule
	\multirow{2}{*}{Error} & \multirow{2}{*}{\% Truncated} & \multirow{2}{*}{\% Censored} & \multicolumn{3}{c}{Number of Internal Knots} \\
        & & & $n=200,$ & $400,$ & $800$ \\
        \midrule
        (a) Standard Normal & 16.1\% & 12.4\% & 1 & 1 & 1 \\
        (b) Standard Extreme Value & 12.3\% & 17.8\% & 1 & 1 & 2 \\
        (c) Mixed Normal & 19.6\% & 18.1\% & 2 & 3 & 4 \\
        (d) Non-Centered Mixed Normal & 20.4\% & 8.9\% & 2 & 2 & 3 \\
        \bottomrule
	\end{tabular}}
	\label{tbl:sim_ops}
\end{table}

Simulations are with different sample sizes of 200, 400, and 800, respectively. Since left-truncation removes data, we keep simulating independent data until accumulated non-truncated data reach the desired sample size, then treat the sample size as a known quantity. For each simulation setup, we generate 1,000 datasets. We use 10 randomly selected initial values as described in Section \ref{sec:likelihood} when fitting each model.

\subsection{Simulation Results}

\begin{table}
\centering
\caption{Results from numerical simulations.}
{	\begin{tabular}{crrrrrrc} 
        \toprule
        Error & Sample Size & Coefficient & Bias $\times 10^3$ & Var$^1 \times 10^3$ & Var$^2 \times 10^3$ & Var$^3 \times 10^3$ & 95\% Coverage \\
        \midrule
        (a) & 200 & $\beta_1$ & 0.1 & 2.3 & 2.8 & 2.5 & 95.5 \\
         & & $\beta_2$ & 7.3 & 23.0 & 26.8 & 25.8 & 95.6 \\
         & 400 & $\beta_1$ & 0.1 & 1.2 & 1.2 & 1.1 & 95.4 \\
         & & $\beta_2$ & 7.2 & 11.8 &  12.4 & 12.8 & 94.2 \\
         & 800 & $\beta_1$ & 0.4 & 0.6 & 0.6 & 0.6 & 94.7 \\
         & & $\beta_2$ & 5.5 & 6.0 & 6.1 & 6.3 & 94.7 \\
         \midrule
        (b) & 200 & $\beta_1$ & 1.7 & 2.4 & 2.6 & 2.7 & 94.9 \\
         & & $\beta_2$ & 11.8 & 23.8 & 26.7 & 26.8 & 94.6 \\
         & 400 & $\beta_1$ & 0.2 & 1.2 & 1.2 & 1.2 & 95.4 \\
         & & $\beta_2$ & 3.4 & 12.1 & 12.8 & 12.2 & 95.9 \\
         & 800 & $\beta_1$ & -0.5 & 0.6 & 0.6 & 0.7 & 94.3 \\
         & & $\beta_2$ & 3.5 & 6.1 & 6.4 & 6.3 & 95.9\\
         \midrule
         (c) & 200 & $\beta_1$ & 3.5 & 8.0 & 9.9 & 9.0 & 95.7  \\
         & & $\beta_2$ & 20.7 & 78.8 & 95.5 & 93.9 & 92.9 \\
         & 400 & $\beta_1$ & -1.6 & 3.3 & 3.8 & 3.7 & 95.2 \\
         & & $\beta_2$ & 2.9 & 33.8 & 38.0 & 40.7 & 94.7 \\
         & 800 & $\beta_1$ & -1.3 & 1.7 & 1.8 & 1.9 & 94.3 \\
         & & $\beta_2$ & 1.4 & 16.9  & 18.2 & 19.5 & 93.5 \\
         \midrule
         (d) & 200 & $\beta_1$ & 0.7 & 1.4 & 1.7 & 1.6 & 95.6  \\
         & & $\beta_2$ & 10.5 & 14.3 & 17.6 & 16.7 & 94.4 \\
         & 400 & $\beta_1$ & 1.0 & 0.7 & 0.8 & 0.8 & 95.8 \\
         & & $\beta_2$ & 7.1 & 7.4 & 7.8 & 9.1 & 92.7 \\
         & 800 & $\beta_1$ & -0.1 & 0.4 & 0.4 & 0.4 & 95.3 \\
         & & $\beta_2$ & -3.2 & 3.7 & 3.9 & 4.0 & 94.4 \\
         \bottomrule
	\end{tabular}}
	\label{tbl:sim_res}
    \begin{tablenotes}
        \item Error distributions: (a) N(0,1), (b) Standard extreme value, (c) $0.5N(0,1) + 0.5N(0, 3^2)$, (d) $0.5N(0,1) + 0.5N(-1, 0.5^2)$.
        \item Var$^1$ is based on the efficient score function estimator in Theorem \ref{them:var_con}, Var$^2$ is based on inverting the information matrix for $\beta$ and $\gamma$, and Var$^3$ is the empirical variance. The coverage probability is obtained using Var$^2$. 
    \end{tablenotes}
\end{table}

We present a comprehensive summary of simulation results in Table \ref{tbl:sim_res}. 
For both $\beta_1$ and $\beta_2$ the biases are negligibly low, although the magnitude of the bias is often larger for $\beta_2$ which likely stems from the lower signal-to-noise ratio for that covariate.

The variance estimates across the three methods of estimation are very similar, especially at larger sample sizes. While the theory presented in this work only guarantees that the variance estimator Var$^1$ obtained from the efficient score function converges to the truth, we see that the variance estimator Var$^2$ obtained by inverting the information matrix of $(\beta, \gamma)$ is close to the value of Var$^1$, and they both are close to  the empirical variance Var$^3$. Further,  all three variances are reduced approximately by half as the sample size is doubled, which is expected from the large sample theory. The confidence intervals created using Var$^2$ provide proper coverages.  The same pattern holds irrespective of error distributions. 

Recall that we use 10 random starting points to fit the model to each dataset, and select the result with the highest log-likelihood value.
Given the known truth, we can examine the convergence of these 10 random starts for each simulated dataset. For both $\beta_1$ and $\beta_2$, we find about $80\sim 95$\% of the randomly selected initial values lead to convergence somewhere close to the truth. 

\begin{figure}
    \centering
    \includegraphics[width=6in]{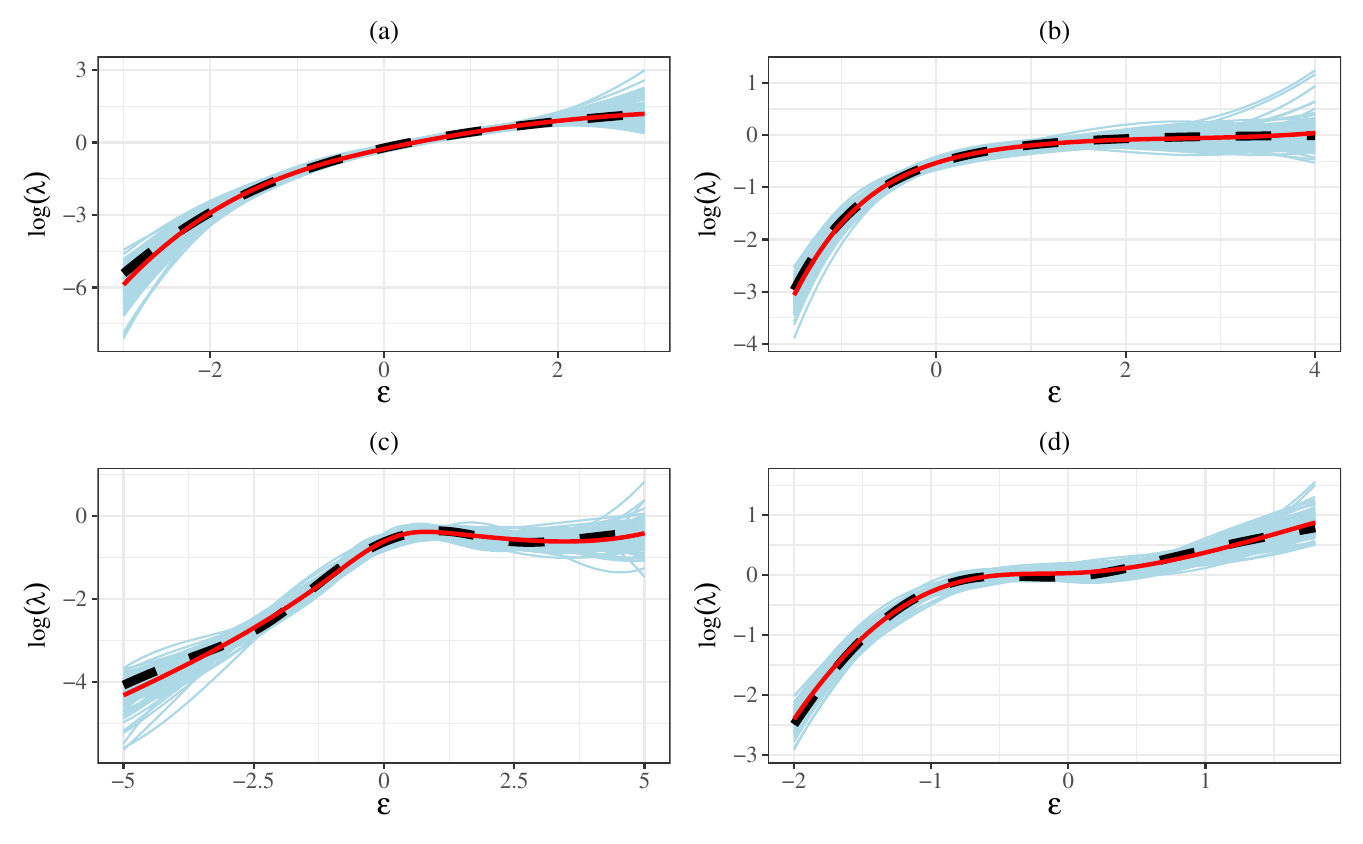}
    \caption{Plots showing the spline estimator for the log hazard function of each error distribution corresponding to Table \ref{tbl:sim_ops}. A sample of individual estimators obtained from 100 datasets are plotted with light-blue solid lines, along with the point-wise average of the estimators obtained from all 1000 datasets plotted with red solid line and the true value plotted with black dashed line. Error distributions: (a) N(0,1), (b) Standard extreme value, (c) $0.5N(0,1) + 0.5N(0, 3^2)$, (d) $0.5N(0,1) + 0.5N(-1, 0.5^2)$.}
    \label{fig:log_haz}
\end{figure}

Along with the regression coefficient estimates, we also examine the estimates of the infinite-dimensional nuisance parameter, the logarithm of the hazard function, in simulations with a sample size of 800.
Figure \ref{fig:log_haz} summarizes the simulation results for each error distribution, including a sample of 100 individual estimates plotted with light-blue solid lines, the point-wise average over all 1,000 datasets plotted with a red solid line, and the ground truth plotted with a black dashed line. The bounds of the $x$-axis represent the middle 95\% of the theoretical error distribution. We see from Figure \ref{fig:log_haz} that the spline estimates of $\log(\lambda_0)$ perform well in simulations.

\section{The 90+ Study of Dementia}
\label{sec:application}

We now apply our proposed method to the dataset of The 90+ Study of Alzheimer's disease and dementia \citep{corrada_dementia_2010}.
This dataset comes from an ongoing study at the Alzheimer's Disease Research Center at the University of California, Irvine.
The study enrolled participants who were at least 90 years old and did not have Alzheimer's disease or dementia at baseline. These individuals were followed via yearly visits.
The main feature of interest is a diagnosis of the cognitive state of each participant, which comes from a panel of experts after reviewing the results of neurologic examination and a battery of
neuropsychological tests.

Understanding dementia onset during lifespan is of great interest. Because dementia may or may not occur during one's lifetime,  a clear description of such a problem involves two components: the lifetime dementia-free probability and the age distribution of dementia onset during one's lifetime if dementia has occurred. We focus on the latter in the analysis of The 90+ Study. Specifically, we are interested in understanding the impact of demographic variables on the age of dementia onset among participants who have developed dementia at some point during their lifetime. For this purpose, we restrict our analysis to 349 available participants with complete information of age at death and age at dementia. 
These participants were from two main cohorts recruited at two different time periods \citep{corrada_dementia_2010, melikyan_recruiting_2019}. Since participants enrolled in the study at different ages that range from 90.1 to 103.0 years old, the data we consider in this analysis are left-truncated. We do not have right-censoring in the dataset because ages at dementia onset are observed for all participants.  Some details of the dataset are given in Table \ref{tbl:90data}, where age at dementia ranges from 90.9 to 109.8 years and age at death ranges from 91.3 to 111.2 years.

\begin{table}
\centering
	\caption{Descriptive statistics of the $349$ participants of The 90+ Study.}
{	\begin{tabular}{lc}
\toprule
        Variable &  Mean (SD) or N (\%)\\
        \midrule
        Enrollment Age & 93.201 (2.745) \\
        Dementia Diagnosis Age & 96.343 (3.222) \\
        Death Age & 98.514 (3.374) \\
        HP$_{90}$ & 0.737 (0.201) \\
        Cohort & \\
        \quad\quad First & 278 (79.7\%) \\
        \quad\quad Second &71 (20.3\%) \\
        Sex & \\
        \quad\quad Female & 251 (71.9\%)\\
        \quad\quad Male & 98 (28.1\%)\\
        Education & \\
        \quad\quad Did Not Graduate College & 208 (59.6\%)\\
        \quad\quad Graduated College & 141 (40.4\%) \\
        Has Had a Stroke? & \\
        \quad\quad No & 211 (60.5\%)\\
        \quad\quad Yes & 138 (39.5\%)\\
        Has Had Heart Disease? & \\
        \quad\quad No & 142 (40.7\%)\\
        \quad\quad Yes & 207 (59.3\%)\\
        \bottomrule
	\end{tabular}}
	\label{tbl:90data}
    \begin{tablenotes}
        \item SD, standard deviation; HP$_{90}$, the proportion of lifespan after 90 years of age that is lived without dementia.
    \end{tablenotes}
\end{table}

To better answer the question of interest, we derive a new variable called the Healthspan Proportion for Age 90, HP$_{90}$, that is the number of healthy years after age 90 divided by the number of total years lived after age 90. Here the number of healthy years refers to the number of years prior to dementia onset. Higher HP$_{90}$ implies better quality of life for an individual in this population of oldest old. We consider the following linear regression model for the logit transformed HP$_{90}$ and leave  the error distribution unspecified:

\begin{eqnarray}
    \log\left(\frac{\text{HP}_{90}}{1 - \text{HP}_{90}}\right) &=& \beta_1(\text{Death Age - 90}) + \beta_2 1(\text{Cohort} = 2) \nonumber \\
    &&\qquad + \ \beta_3 1(\text{Sex = Female}) \label{eqn:90mod} \\
    &&\qquad + \ \beta_4 1(\text{Education = Graduated College})  \nonumber\\
    &&\qquad + \ \beta_5 1(\text{Stroke = Yes or Heart Disease = Yes})  + \text{Error}. \nonumber
\end{eqnarray}

Since HP$_{90}$ potentially ranges from 0 to 1, the logit transformed HP$_{90}$ can take any real value, and it is left-truncated because age at dementia is left-truncated.  
We apply our proposed method to fit a left-truncated regression model to the data. 
We use cross-validation to select the number of internal knots for the considered B-spline basis expansion, and ultimately choose two internal knots.
From the regression results given in Table \ref{tbl:reg_model} we see that increased longevity is associated with higher HP$_{90}$, and the second cohort has higher HP$_{90}$ than the first cohort. 

\begin{table}
\centering
	\caption{Results of the sieve method for fitting model (\ref{eqn:90mod}) to the left-truncated data of The 90+ Study.}
{	\begin{tabular}{lccc}
\toprule
        Variable &  Estimate & 95\% CI & P-Value\\
        \midrule
        Death Age & 0.0578 & (0.0221, 0.936) & 0.0015 \\
        Cohort & \\
        \quad\quad First & Referent & -- & -- \\
        \quad\quad Second & 0.3312 & (0.0158, 0.6465) & 0.0396 \\
        Sex & \\
        \quad\quad Male & Referent & -- & --\\
        \quad\quad Female & -0.0660 & (-0.3431, 0.2109) & 0.6403\\
        Education & \\
        \quad\quad Did Not Graduate College & Referent & -- & -- \\
        \quad\quad Graduated College & 0.1638 & (-0.0882, 0.4157) & 0.2026 \\
        Comorbidity: Stroke or CHD & \\
        \quad\quad No & Referent & -- & --\\
        \quad\quad Yes & 0.0200 & (-0.3332, 0.3733) & 0.9116\\
        \bottomrule
	\end{tabular}}
	\label{tbl:reg_model}
\end{table}

\section{Discussion}
\label{sec:discuss}

Left-truncation and right-censoring are often present in prospective cohort studies. Although linear regression models with unspecified error distributions are appealing due to  simplicity, their applications to censored or truncated event time data are rather limited, mainly because of lacking easily implementable estimating methods with desirable statistical properties. In this work we show that the semiparametric sieve maximum likelihood approach for bundled parameters possesses nice properties in dealing with left-truncated and right-censored data in linear regression models. The method yields asymptotically efficient and normally distributed estimates for regression coefficients, with estimators obtained using an easily implementable gradient search algorithm. We expect this work would encourage the proper use of linear regression models in analyzing event time data that are subject to left-truncation and right-censoring.



\bmsection*{Acknowledgments}

We are grateful to Dr. Maria Corrada for sharing The 90+ Study dataset.

\bmsection*{Financial disclosure}

The work is supported in part by NIH grant RF1 AG075107 and UCI Alzheimer's Disease Research Center Grant P30 AG066519, and 
by NSF grant DMS 2412746.

\bmsection*{Conflict of interest}

The authors declare no potential conflict of interests.

\bibliography{references}

\bmsection*{Supporting information}

Readers can access the \texttt{R} code used for the numerical implementations in this article on the first author's GitHub page \texttt{https://github.com/srmatth/ltrc}.

\appendix

\bmsection{Likelihood Function}
\vspace*{12pt}
Let $\bar{F}_A(a)$ denote the survival function of an arbitrary random variable $A$.
We proceed with looking at two sub-distributions that correspond to $\Delta = 1$ and $\Delta = 0$, respectively. We have
\begin{align}
    &{\rm pr}(Y \leq y, \Delta = \delta \mid Y > T, T=t, X=x) \nonumber \\
    & \qquad = \frac{{\rm pr}(Y \leq y, \Delta = \delta, Y > t \mid T=t,X=x)}{{\rm pr}(Y > t \mid T=t, X=x)}1(y>t).
    \label{eqn:a1}
\end{align}

From our assumption, $Y^*$ is independent of $(C,T)$  given $X$. Consider first the denominator in (\ref{eqn:a1}). We have
\begin{align}
{\rm pr}(Y > t \mid T=t,X=x) &= {\rm pr}(Y^* > t, C>t \mid T=t,X=x)\nonumber \\
& = \bar{F}_{Y^* \mid X}(t \mid x) \bar{F}_{C\mid T,X}(t\mid t,x).
\label{eqn:denom}
\end{align}

Now consider the numerator in (\ref{eqn:a1}). For the case where $\Delta = 1$, we have
\begin{align*}
    &{\rm pr}(Y \leq y, \Delta = 1, Y > t \mid T=t,X=x) \\
    &\qquad = {\rm pr}(Y^* \leq y, Y^* < C, Y^* > t \mid T=t, X=x) \\
    &\qquad = \int_t^y \int_u^\infty f_{Y^*, C \mid T, X}(u,v \mid t,x)dvdu \\
    &\qquad = \int_t^y \int_u^\infty f_{Y^*\mid X}(u \mid x)f_{C\mid T, X}(v\mid t,x) dvdu \\
    &\qquad = \int_t^y f_{Y^* \mid X}(u \mid x) \bar{F}_{C\mid T,X}(u \mid t,x)du.
\end{align*}
When $\Delta = 0$ we have
\begin{align*}
    &{\rm pr}(Y \leq y, \Delta = 0, Y > t \mid T=t,X=x) \\
    &\qquad = {\rm pr}(C \leq y, Y^* > C, C > t \mid T=t,X=x) \\
    &\qquad = \int_t^y \int_v^\infty f_{Y^*, C \mid T, X}(u,v \mid t,x)dudv \\
    &\qquad = \int_t^y \int_v^\infty f_{Y^*\mid X}(u \mid x)f_{C\mid T,X}(v\mid t,x) dudv \\
    &\qquad = \int_t^y f_{C \mid T, X}(v \mid t,x) \bar{F}_{Y^*\mid X}(v \mid x)dv.
\end{align*}

Thus we obtain the following conditional density function:
\begin{align*}
&f_{Y,\Delta \mid Y>T, T,X}(y,\delta \mid y>t,  t,x) \\
& \qquad = \frac{\left\{f_{Y^* \mid X}(y \mid x) \bar{F}_{C\mid T,X}(y \mid t,x)\right\}^\delta\left\{f_{C \mid T, X}(y \mid t, x) \bar{F}_{Y^*\mid X}(y \mid x)\right\}^{1-\delta}}{\bar{F}_{Y^* \mid X}(t \mid x) \bar{F}_{C\mid T,X}(t \mid t, x)} {1}(y > t).
\end{align*}
Then the joint density function of left-truncated data is
\begin{align}
    &f_{Y,\Delta, T, X \mid Y>T}(y,\delta,t,x \mid y>t) {1}(y>t) \notag \\
    &\qquad =f_{Y,\Delta \mid T,X,Y>T}(y,\delta \mid t,x,y>t) f_{T,X\mid Y>T}(t, x \mid y>t) {1}(y>t) \notag \\
    &\qquad = \frac{\left\{f_{Y^* \mid X}(y \mid x) \bar{F}_{C\mid T,X}(y \mid t, x)\right\}^\delta\left\{f_{C \mid T,X}(y \mid t, x) \bar{F}_{Y^*\mid X}(y \mid x)\right\}^{1-\delta}}{\bar{F}_{Y^* \mid X}(t \mid x) \bar{F}_{C\mid T,X}(t\mid t,x)} \notag \\
    &\qquad \qquad \times f_{T,X\mid Y>T}(t, x \mid y>t)  {1}(y > t).
    \label{eqn:app_lklhd}
\end{align}
Dropping the factors that are free of $(\beta, \lambda_e)$ from (\ref{eqn:app_lklhd}), we obtain the likelihood:
\begin{align*}
    L(\beta, \lambda_e \mid Y,\Delta,T,X) &= \frac{\left\{f_{Y^* \mid X}(Y \mid X) \right\}^\Delta\left\{ \bar{F}_{Y^*\mid X}(Y \mid X)\right\}^{1-\Delta}}{\bar{F}_{Y^* \mid X}(T \mid X)}.
\end{align*}
Using hazard functions, we can rewrite the above likelihood function as the following:
\begin{align*}
    L(\beta, \lambda_e \mid Y,\Delta,T,X) &= \frac{\left\{\lambda_{Y^* \mid X}(Y \mid X)\right\}^\Delta \exp\left\{-\int_{-\infty}^Y \lambda_{Y^* \mid X}(u\mid X)du \right\}}{\exp\left\{-\int_{-\infty}^T \lambda_{Y^* \mid X}(u\mid X)du\right\}} \\
    &= \left\{\lambda_{Y^* \mid X}(Y \mid X)\right\}^\Delta \exp\left\{-\int_{T}^Y \lambda_{Y^* \mid X}(u\mid X)du\right\} \\
    &= \left\{\lambda_{e}(Y - X^\T\beta)\right\}^\Delta \exp\left\{-\int_{T - X^\T\beta}^{Y - X^\T\beta} \lambda_{e}(u)du\right\}.
\end{align*}







\setcounter{table}{0}
\renewcommand{\thetable}{\arabic{table}}

\setcounter{figure}{0}
\renewcommand{\thefigure}{\arabic{figure}}

\end{document}